\documentclass[openacc]{rstransa} 
\usepackage[utf8]{inputenc}
\usepackage{amsmath, amssymb}
\usepackage{graphicx}
\usepackage{caption}
\usepackage{times}
\usepackage{orcidlink}
\usepackage[square,numbers]{natbib}
\usepackage{amssymb}

\usepackage{subcaption}

\titlehead{Research}
\begin{document}

\title{Recurrent Coronal Jets and QPPs: Periodic Reconnection and Localized Heating Across Quiet-Sun to Active Regions}

\author{
Sudheer K. Mishra\orcidlink{0000-0003-2129-5728}$^{1}$,
Kartika Sangal\orcidlink{0000-0003-0596-5277}$^{2}$,
Balveer Singh\orcidlink{0000-0001-6234-6400}$^{3}$,
Ayumi Asai\orcidlink{0000-0002-5279-686X}$^{1}$,
A. K. Srivastava\orcidlink{0000-0002-1641-1539}$^{4}$,
and Ding Yuan\orcidlink{0000-0002-9514-6402}$^{2}$}

\address{$^{1}$Astronomical Observatory, Kyoto University, Sakyo, Kyoto 606-8502, Japan\\
$^{2}${Key Laboratory of Solar Activity and Space Weather, School of Aerospace, Harbin Institute of Technology, Shenzhen, People's Republic of China}\\
$^{3}$Division of Mathematics, School of Science and Engineering, University of Dundee, DD1 4HN, UK\\
$^{4}$Department of Physics, Indian Institute of Technology (BHU), Varanasi-221005, India}

\subject{Solar and Heliospherics}

\keywords{Solar Corona, Solar Flares, Coronal Jets, Quasi-Periodic Pulsations (QPPs), Coronal Heating, Magnetic Reconnection, Fan–Spine Topology}

\corres{Sudheer K. Mishra\\
\email{sudheer1018@gmail.com}
\email{mishra.sudheer.6r@kyoto-u.ac.jp}}

\begin{abstract}
We analyze quasi-periodic pulsations (QPPs) in recurrent coronal jets driven by periodic magnetic reconnection associated with successive flux emergence in fan-spine magnetic topologies. Using {\it Atmospheric Imaging Assembly} (AIA) onboard the {\it Solar Dynamics Observatory} (SDO), we investigate three long-lived recurring jets spanning quiet-Sun to moderate-field-strength regions, each exhibiting recurrent eruptions linked to episodic reconnection. Wavelet analysis of multithermal AIA EUV jet-base light curves detects QPPs with periods of 6-13~min, exceeding typical $p$-mode oscillation periods. Distance-time analysis reveals quasi-periodic propagating ridges, interpreted as recurrent field-aligned plasma ejections, and morphological similarity to slow magnetoacoustic waves, which cannot be entirely excluded. However, the dominant photospheric unsigned flux periodicities of 10-32~min at the jet source regions favor the reconnection-driven interpretation. DEM analysis confirms multithermal plasma with the hottest emission concentrated near the jet base, and the QPP periods fall well below both radiative and conductive cooling timescales, implying persistent localized heating within the fan-spine configuration. These results demonstrate that periodic reconnection in fan-spine topologies drives recurrent jet eruptions and contributes to localized coronal heating across quiet Sun, moderate-field-strength regions, and active regions.
\end{abstract}

\maketitle
\section{Introduction}
Solar jets are collimated plasma ejections observed in active, quiet, and coronal hole regions at diverse spatio-temporal scales. These jets are widely regarded as key contributors to the mass and energy transport through various coupled atmospheric layers of the Sun \cite{1992PASJ...44L.173S, 2016SSRv..201....1R}. Since the discovery of coronal and X-ray jets, it has been suggested that they may play an important role in coronal heating \cite{1983ApJ...272..329B, 1982ApJ...261..375K}. The initiation of these jets is primarily governed by the flux emergence and, increasingly, flux cancellation, where newly emerging or canceling opposite-polarity patches reconnect with the pre-existing coronal field to launch magnetically guided jets \cite{1992PASJ...44L.173S, 2016ApJ...822L..23P}. In addition to flux emergence and cancellation, coronal jets may also be triggered by kink instability developing within the fan-spine magnetic topology \cite{2009ApJ...691...61P} or via the magnetic breakout mechanism, wherein reconnection between the energized inner closed flux system and the overlying field drives the impulsive jet outflow \cite{2017Natur.544..452W, 2018ApJ...852...98W}.
More broadly, magnetic reconnection governs jet formation, with its temperature, morphology, and dynamics strongly dependent on the height of the reconnection site in the solar atmosphere \cite{1995Natur.375...42Y, 2013PASJ...65...62T}. 
Coronal and chromospheric null points, quasi-separatrix layers (QSLs), and current sheets are recognized as preferred sites for efficient magnetic energy release, making them prime regions for plasma heating and potentially contributing to coronal heating \cite{2009ApJ...704..485T, 2013A&A...555A..19G}. 
When reconnection occurs at these sites, 
triggered by episodic or successive flux emergence or modulated by MHD waves, it can drive recurrent jet activity \cite{2012A&A...542A..70M, 2013A&A...555A..19G, 2023ApJ...945..113M, 2024NatAs...8..706L}. The associated multi-thermal plasma releases energy periodically or quasi-periodically, producing observable signatures of quasi-periodic pulsations (QPPs), consistent with models of oscillatory reconnection and wave-reconnection coupling \cite{2012A&A...548A..98M, 2017ApJ...844....2T}.\\

 Periodic magnetic reconnection, or reconnection modulated by magnetohydrodynamic (MHD) waves, is widely regarded as one of the key mechanisms for generating QPPs in solar and stellar flares \cite{2009SSRv..149..119N, 2021SSRv..217...66Z}. While QPPs are commonly observed in solar flares, reports of QPPs in small-scale ejections, such as coronal jets, remain relatively scarce. Recent high-resolution imaging observations from AIA onboard SDO and spectroscopic observations from IRIS have revealed QPPs signatures in blowout jets, likely driven by periodic magnetic reconnection or p-mode modulation \cite{2012A&A...542A..70M, 2022FrASS...932099L, 2022ApJ...933...21K, 2023ApJ...945..113M, 2024SoPh..299...88K}.  Theoretical models and numerical simulations have increasingly supported the idea that both wave- and reconnection-driven mechanisms coexist or interact, leading to complex QPP signatures \cite{2010ApJ...714.1762P, 2021MNRAS.505...50G, 2022MNRAS.511.4134S, 2024RSPTA.38230220S}. Recent ultra-high resolution imaging observations of EUI onboard Solar Orbiter have shown the persistent coronal null for over an hour, launching repetitive jets consisting of QPP-like signatures from a fan-spine topology \cite{2023NatCo..14.2107C}. Radiative transfer simulation further suggests that 
in active regions, repetitive reconnection within fan–spine configurations can drive periodic jets that form super-hot loops \citep{2024NatAs...8..706L}.
Most QPP detections reported so far are associated with active regions or other sites of strong magnetic fields. In contrast, in quiet-Sun and moderate field strength regions, jet activity typically appears as a compact bright base, commonly identified as coronal bright points (CBPs) or X-ray bright points (XBPs), accompanied by an elongated spine that channels plasma into the overlying atmosphere \cite{1992PASJ...44L.173S, 1999SoPh..190..167A, 2019LRSP...16....2M}.\\

The oscillatory nature of coronal bright points (CBPs) was first reported by \cite{1979SoPh...63..119S} and later confirmed by multi-instrument observations, which revealed characteristic periods ranging from approximately 4 to 60 minutes \cite{2003A&A...398..775M, 2004A&A...418..313U, 2008A&A...485..289K}. High-resolution imaging and spectroscopic observations further indicate that the quasi-periodic pulsations (QPPs) observed in these systems are linked to regimes of periodic magnetic reconnection and reconnection modulated by magnetohydrodynamic (MHD) waves \cite{2008A&A...489..741T, 2015ApJ...810..163C, 2015ApJ...806..172S}. These CBPs associated with jet activity frequently exhibit recurrent plasma ejections along an elongated spine, accompanied by impulsive brightenings near the jet bases. These brightenings often display oscillatory or quasi-oscillatory behavior, 
generally interpreted as signatures of periodic magnetic reconnection or reconnection modulated by MHD waves. Statistical analyses of CBP-related jets reveal that quasi-periodic intensity enhancements commonly precede the onset of coronal jets, likely driven by MHD oscillation or wave-like motions that trigger episodic reconnection \cite{2010ApJ...710.1806D,2018ApJ...855L..21B}. Extensive numerical simulations further demonstrate that, in the presence of fan-spine magnetic configurations, CBPs can launch quasi-periodic jets associated with repeated episodes of magnetic flux emergence and cancellation \cite{2021MNRAS.505...50G, 2023ApJ...958L..38N, 2024NatAs...8..706L}. \\

In addition, spectral and differential emission measure (DEM) analyses suggest that CBP plasmas exhibit a cool component (log $T \approx 5.2$) along with dominant coronal components (log $T \approx 6.2$–6.3) \cite{2010ApJ...710.1806D, 2013ApJ...768...32C}.  Recent high-resolution observations and modeling suggest that a coronal null embedded within a fan–spine topology can host super-hot plasma (in the presence of a coronal null exceeding 10~MK) both near the jet base and in the vicinity of the null point for several hours \cite{1996PASJ...48..353Y, 2023NatCo..14.2107C}. In such configurations, long-lived CBPs and their associated jet activity are driven by successive episodes of flux emergence and cancellation and are natural candidates for sustaining periodic magnetic reconnection. These long-lived, highly energized magnetic structures are therefore strong candidates for contributing to localized coronal heating. \\

In this study, we examine three long-lived jets associated with CBPs located in regions of different magnetic field strengths: Jet1 in a quiet-Sun region, Jet2 in a moderate-field region, and Jet3 in an active region of the Sun. These jets typically exhibit inverted-Y (anemone) structures and display quasi-periodic intensity enhancements in multiple AIA channels. Characteristic periods range from about 6 to 13 minutes, depending on the different AIA EUV wavelengths. We investigate the magnetic origin of these jets, along with their thermal properties, kinematics, oscillatory behavior, and QPP signatures, and assess their potential contribution to steady, localized coronal heating. The present paper is organized as follows. We describe the observational data and analysis in section \ref{sec:data}. Section \ref{sec: results} elucidates the observational results and the explanation of QPPs in the coronal jets. Section \ref{sec: conclusion} presents the discussions, the concluding remarks, and future perspectives for further exploration.

\section{Observational Data and Analysis}
\label{sec:data}
The Atmospheric Imaging Assembly (AIA) onboard SDO observes the solar atmosphere in ten EUV and UV channels with 1.5'' resolution, a 0.6'' pixel scale, and a 12~s cadence \cite{2012SoPh..275...17L, 2012SoPh..275....3P}. We use Level~1.5 data processed with \textit{aia\_prep}. Among the EUV channels (94, 131, 171, 193, 211, 304, and 335~{\AA}), the 304~{\AA} band is excluded from the DEM analysis because it is optically thick. To investigate the thermal properties of plasma energized by periodic reconnection in the fan-spine-like configuration, we apply the sparse-inversion DEM method \cite{2015ApJ...807..143C} using the six optically thin EUV channels. The temperature range is set to $\log T = 5.5$--$7.0$ with a bin size of $\Delta\log T=0.1$. The DEM maps indicate that the jet base contains hotter plasma, whereas the extended spine shows coronal-temperature emission.

To investigate the periodic or quasi-periodic intensity enhancements at the jet base and magnetic source region, we utilize wavelet analysis \cite{1998BAMS...79...61T}. We extracted the light curves of different AIA channels and total unsigned flux from the LOS HMI magnetogram. Before applying wavelet analysis, we detrend the time series and remove long-term variations. We applied a running average with a boxcar window of $40$ data points, corresponding to $\approx$~8~minutes for AIA channels and $\approx$~30~minutes for the HMI LOS magnetogram. The detrended signal was then obtained by subtracting this smoothed trend from the original time series, following the standard approach used in QPP studies \citep{1998BAMS...79...61T}. For wavelet analysis, we used the Morlet function as the mother function \cite{1998BAMS...79...61T}. The cross-hatched region is referred to as the cone of influence (COI), and the power within this region is not considered significant due to the contours, and the corresponding periods are considered edge effects. The overplotted red contours represent the 95\% significance level. We estimated these levels using a red-noise model \cite{1998BAMS...79...61T}. The power lies within these significant contours, and the corresponding periods are considered significant. To compute the dominant period in the light curve, we used the global wavelet spectrum, defined as the time-averaged wavelet spectrum. The detailed analysis and results are discussed in the sections below.

\section{Observational Results}
\label{sec: results}
We present multi-wavelength imaging observations of three long-lived, CBP-like jets that occur in fan–spine magnetic configurations with differing magnetic field strengths. Morphologically, a fan-spine topology is a 3-dimensional reconnection geometry, defined as a dome-shaped magnetic structure in which a compact polarity patch is surrounded by an opposite-polarity field, forming a coronal null point with an associated fan surface and spine field lines. Jet1 and Jet2 appeared on 2 October 2024 in a quiet-Sun region and in a region of moderate field strength, respectively, whereas Jet3 was observed on 1 June 2023 in an evolving moderate field active region of the Sun. The magnetic topology and triggering mechanism of the eruptive phase of Jet3 (a subset of the full observations) have been analyzed by \cite{2024ApJ...962L..38D, 2024cosp...45.1960M}, while its long-lived activities, from small-scale anemone jets to large-scale homologous recurring jet eruptions, overall triggering mechanism, thermal response, and contribution to localized coronal heating are discussed in \cite[submitted;][]{Mishra2026submitted}. We investigate QPPs linked to periodic magnetic reconnection in recurrent coronal jets, with AIA morphology suggesting a fan-spine-like reconnection geometry. We show this reconnection scenario supports sustained, repeated energy release and localized heating, maintaining hot coronal emission for hours across quiet-sun to active-region environments. We inspected the e-Callisto network and NOAA/SWPC solar radio event catalogs for all five events and found no associated radio burst activity, consistent with the small spatial scales (a few Mm) of CBP-driven jets, where non-thermal electron beams, if present, are unlikely to produce detectable radio signatures.

\subsection{Multi-wavelength Observations of Anemone Jets in the Quiet Sun}
Panel (a) of Figure~1 shows the HMI line-of-sight (LOS) magnetogram of the quiet-Sun anemone jet source region, revealing mixed-polarity fields with a central negative-polarity patch surrounded by positive polarity, a configuration consistent with a fan-spine magnetic topology, favorable for jet triggering via interchange reconnection. Panels (b)-(d) show the co-temporal appearance of the jet at $t \approx$ 05:00~UT in AIA 304~\AA\ (cool chromospheric/transition-region plasma, $\sim$0.05~MK), 171~\AA\ (warm coronal plasma, $\sim$0.6--0.8~MK), and 131~\AA\ (cool and hot flare plasma, 0.4 MK and $\sim$10~MK; Fe~{\sc xxi}), revealing its multi-thermal nature. The fan-spine reconnection ejects cool, warm, and hot plasma components simultaneously along open spine field lines, confirming the multi-thermal character of the anemone jet.\\
To investigate the quasi-periodic flux emergence and cancellation within the mixed-polarity region, we defined a yellow dashed box in panels (a) to extract the temporal evolution of the unsigned LOS magnetic flux. Panel (e) shows the normalized HMI unsigned flux of Jet1 (black curve), which decreases from 1.0 to $\sim$0.4 over $\sim$540~minutes, overlaid with a 1800s (40 data points) running mean (red) used for detrending. Panel (f) shows the resulting detrended flux, revealing successive, quasi-periodic episodes of flux emergence and cancellation superimposed on the long-term decay. Panel (g) displays the Morlet wavelet power spectrum of the detrended flux, where white contours enclose regions of statistically significant power exceeding the 95$\%$ red-noise significance level (lag-1 autocorrelation $a = 0.44$); the hatched region marks the cone of influence (COI), where edge effects reduce reliability. Panel (h) displays the extracted global wavelet power spectrum (GWS), identifying a dominant periodicity of $\sim$22.74~min (red dashed line) that significantly exceeds the 95\% red-noise background (blue dashed line, $a = 0.44$), indicating a persistent quasi-periodic flux emergence/cancellation, which is responsible for the recurrent jet activity. \\
\begin{figure*}[ht]
\centering
\includegraphics[width=12.0cm, height=2.5cm]{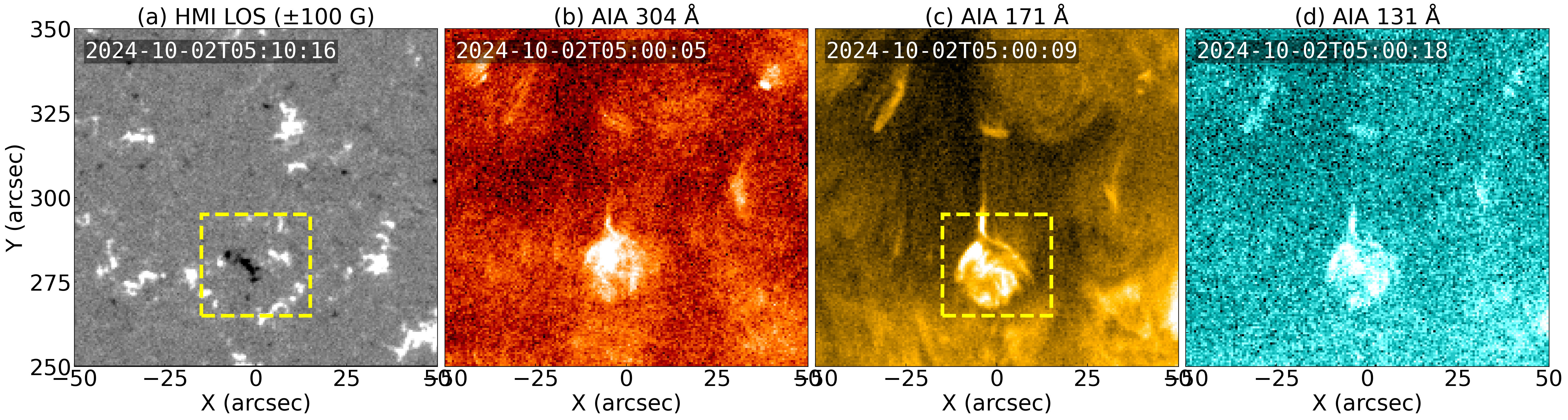}
\includegraphics[width=12.0cm, height=2.5cm]{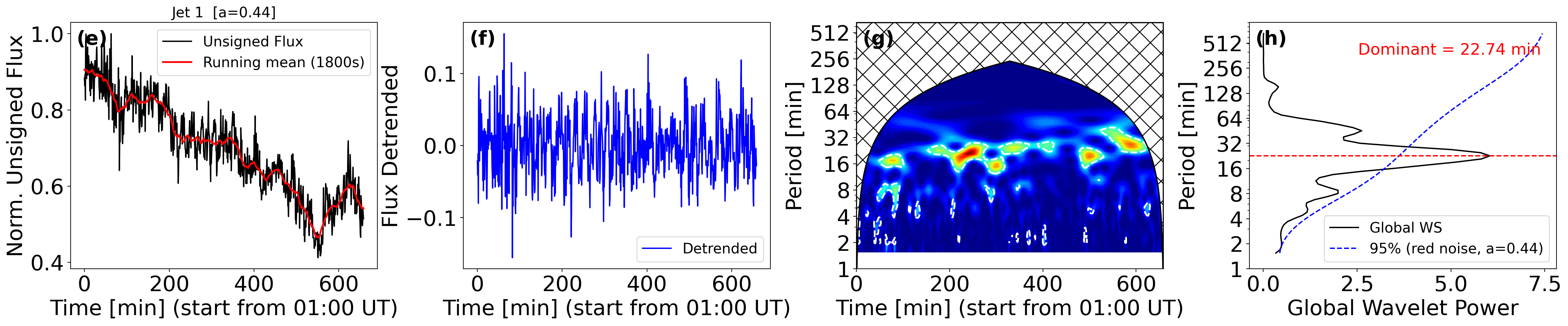} 
\mbox{
\includegraphics[width=3.0cm, height=4.cm,angle=90]{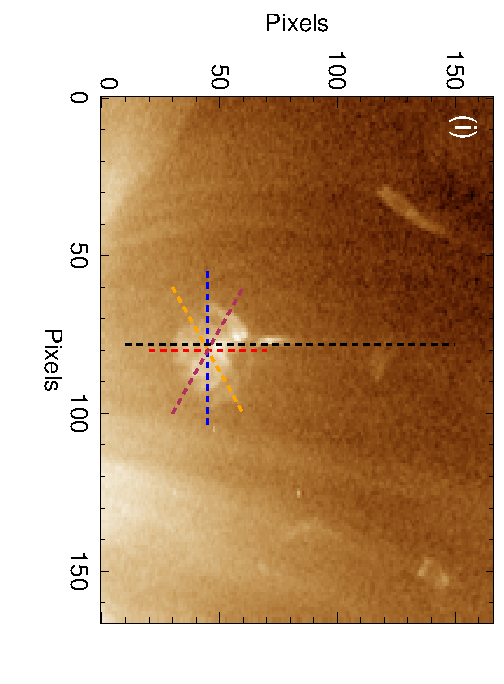}
\hspace{-0.4cm}
\includegraphics[width=3.0cm, height=4.cm, angle=90]{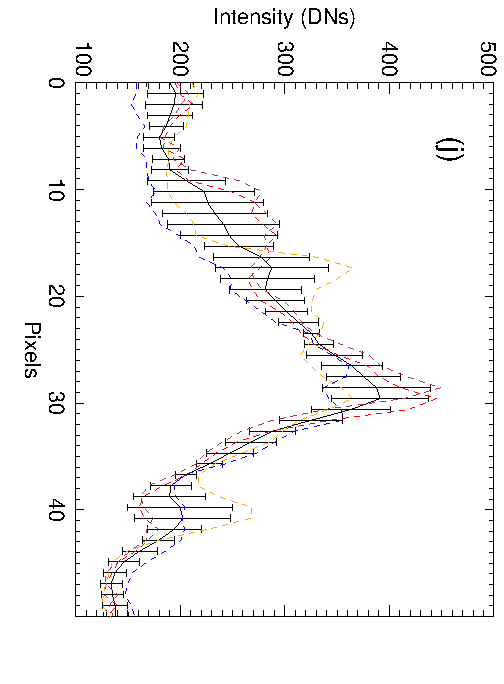} 
\hspace{-0.4cm}
\includegraphics[width=4.0cm, height=3.cm]{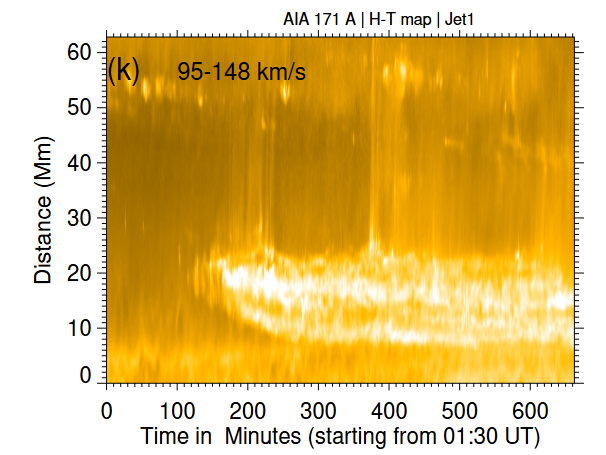} 
}
\caption{Multi-wavelength observations of Jet~1 from the quiet-Sun region. (a) SDO/HMI LOS magnetogram ($\pm$100~G) showing the mixed-polarity anemone source region; (b) AIA~304~\AA, (c) 171~\AA, and (d) 131~\AA\ images at $t \approx$ 05:00~UT, revealing the multi-thermal jet structure. Yellow dashed boxes in panels (a) and (c) mark the ROIs used for unsigned flux and light-curve extraction from the jet base, respectively. (e) Normalized unsigned LOS flux (black) with a 1800~s running mean (red) used for detrending; (f) detrended flux showing quasi-periodic flux emergence and cancellation; (g) Morlet wavelet power spectrum with white contours marking the 95\% red-noise significance level ($a=0.44$) and hatching indicating the cone of influence (COI); (h) global wavelet power spectrum (black) with the 95\% red-noise background (blue dashed) identifying a dominant period of $\sim$22.74~min (red dashed line). (i) AIA~171~\AA\ image with the jet-spine slit (dashed black) and four cross-slits (colored) centered on the jet base; (j) mean intensity variation profiles along each cross-slit (colored curves) and their pixel-wise average (black, $\pm$1$\sigma$) used to estimate the jet-base diameter; (k) AIA~171~\AA\ height-time map revealing recurrent propagating ridges with the projected plane-of-sky velocities of 95-148 km s$^{-1}$, confirming periodic fast coronal
outflows driven by fan-spine reconnection.}
    \label{fig:fig1}
\end{figure*}
To characterize the jet-base geometry and estimate the loop length for Jet~1, we placed four artificial slits, horizontal, vertical, and two diagonal, centered on the jet base, plus an extended slit along the jet spine (panel~i of Figure~1). The pixel-wise mean intensity profile (black curve, panel~j), averaged across all four orientations, with error
bars denoting the model-independent standard deviation, provides a geometry-averaged estimate of the jet-base spatial extent without assuming a functional form for the intensity distribution. The average intensity variation along the slit yields the jet-base diameter; adopting the standard hemispherical approximation for coronal bright points
\citep{2019LRSP...16....2M} as a first-order model, since the line-of-sight depth cannot be constrained from AIA imaging alone, the loop length is then estimated as $L = \pi R$, where $R$ is the loop-base radius. The derived quantities should be treated as upper limits under the assumption of a unity filling factor. The estimated loop length for Jet~1 is listed in Table~1. The AIA~171~\AA\ height-time map (panel~k) reveals bright inclined ridges propagating outward from ${\sim}10$-20~Mm with projected plane-of-sky velocities of 95--148~km~s$^{-1}$, confirming fast coronal outflows driven by fan-spine reconnection. The first recurrent jet episode is discernible at $t \approx$ 03:10~UT, and the continue at $t \approx$ 13:00~UT, spanning ${\sim}$660~min of activity that closely overlaps with the quasi-periodic flux emergence and cancellation interval identified in panel~(e), directly linking the recurrent jet driving to the underlying photospheric flux evolution.

\subsection{Multi-wavelength Observations of Recurring Jets from a Moderate-Field-Strength Region}
Panel (a) of Figure~2 shows the HMI LOS magnetogram ($\pm$200~G) of the Jet~2 source region in a moderate-field-strength region in contrast to the quiet-Sun environment of Jet~1, showing stronger, spatially extended mixed-polarity flux concentrations consistent with a more
energetic reconnection environment. Panels (b-d; Figure~2) show the co-temporal multi-thermal structure of Jet~2 at $t \approx$ 03:10~UT in AIA 304~\AA\, 171~\AA\, and 131~\AA\, with a collimated jet and bright base visible across all channels, confirming multi-thermal fan-spine reconnection.\\
Panel~(e) of Figure~2 shows the normalized unsigned LOS flux extracted from the yellow dashed boxes in panels (a). Unlike the monotonic decline of Jet~1, the flux rises from $\sim$0.5 to $\sim$1.0 during the first $\sim$260~min before and after that start declining, reflecting active flux emergence followed by cancellation. Panel~(f) shows the detrended flux (1800~s running mean removed), revealing intermittent quasi-periodic fluctuations. Panel~(g) of Figure~2 presents the Morlet wavelet power spectrum with significant power (white contours, 95\% red-noise level, $a = 0.32$) concentrated at $\sim$2-4~min and $\sim$22-32~min during the late-phase cancellation. Panel (h) of Figure~2 displays the global wavelet spectrum, extracting a dominant period of $\sim$22.74~min (red dashed) exceeding the 95\% red-noise background (blue dashed, $a = 0.32$), which is consistent with Jet~1, suggesting a common quasi-periodic flux emergence driver independent of the source region field strength.\\
To investigate the kinematics and loop geometry of Jet~2, we followed the same slit-based approach as Jet~1 (panels~i-j) to estimate the
jet-base spatial extent and loop length $L = \pi R$. The estimated loop length for Jet2 is mentioned in Table~1. The AIA~171~\AA\ height-time map is extracted from a black slanted slit (panel~i; Figure~2) and displayed in panel~k, showing bright propagating ridges extending from ${\sim}$10-25~Mm (expanding fan-plane) with projected plane-of-sky velocities of 168-215~km~s$^{-1}$, which is faster than Jet~1 (95-148~km~s$^{-1}$), consistent with the stronger background field driving more energetic reconnection outflows. The first jet episode appears at $t \approx$ 02:10~UT and the last at $t \approx$ 07:20~UT, spanning ${\sim}$300~minutes, a window that closely overlaps the active flux emergence and cancellation phase in panel~(e), directly linking the recurrent jet activity to the underlying quasi-periodic photospheric
flux evolution.
\begin{figure*}[ht!]
\centering{}
\includegraphics[width=12.0cm, height=2.5cm]{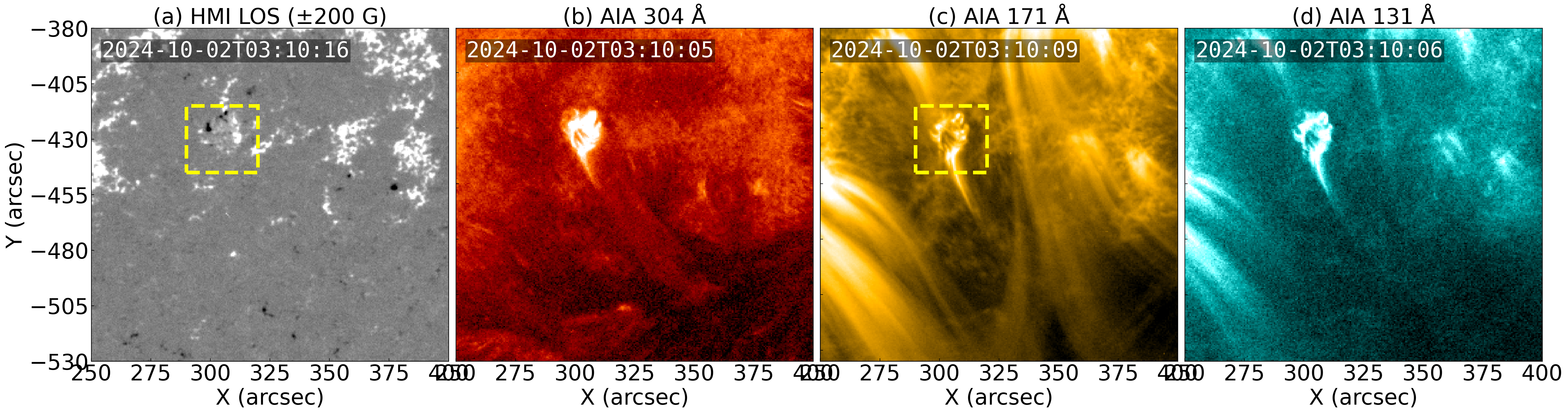}
\includegraphics[width=12.0cm, height=2.5cm]{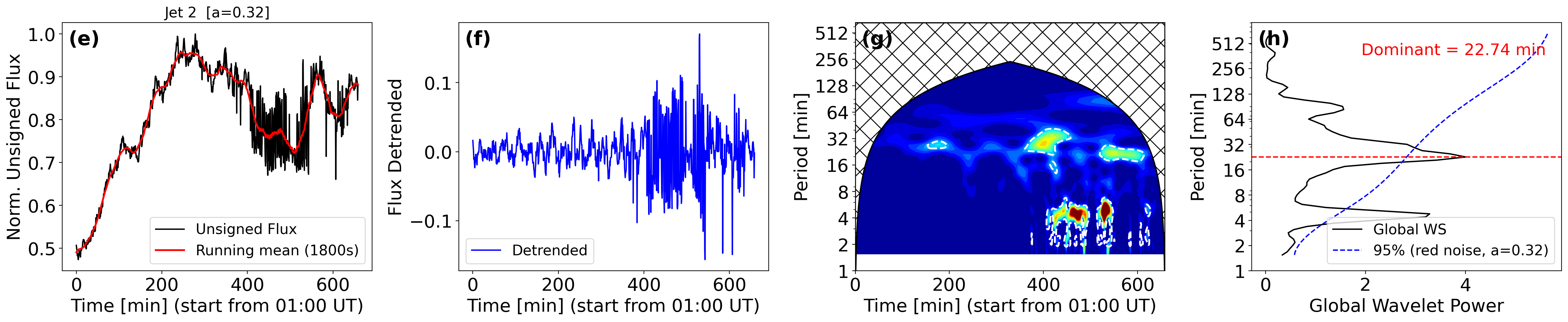} 

\mbox{
    \includegraphics[width=3.cm, height=4.cm,angle=90]{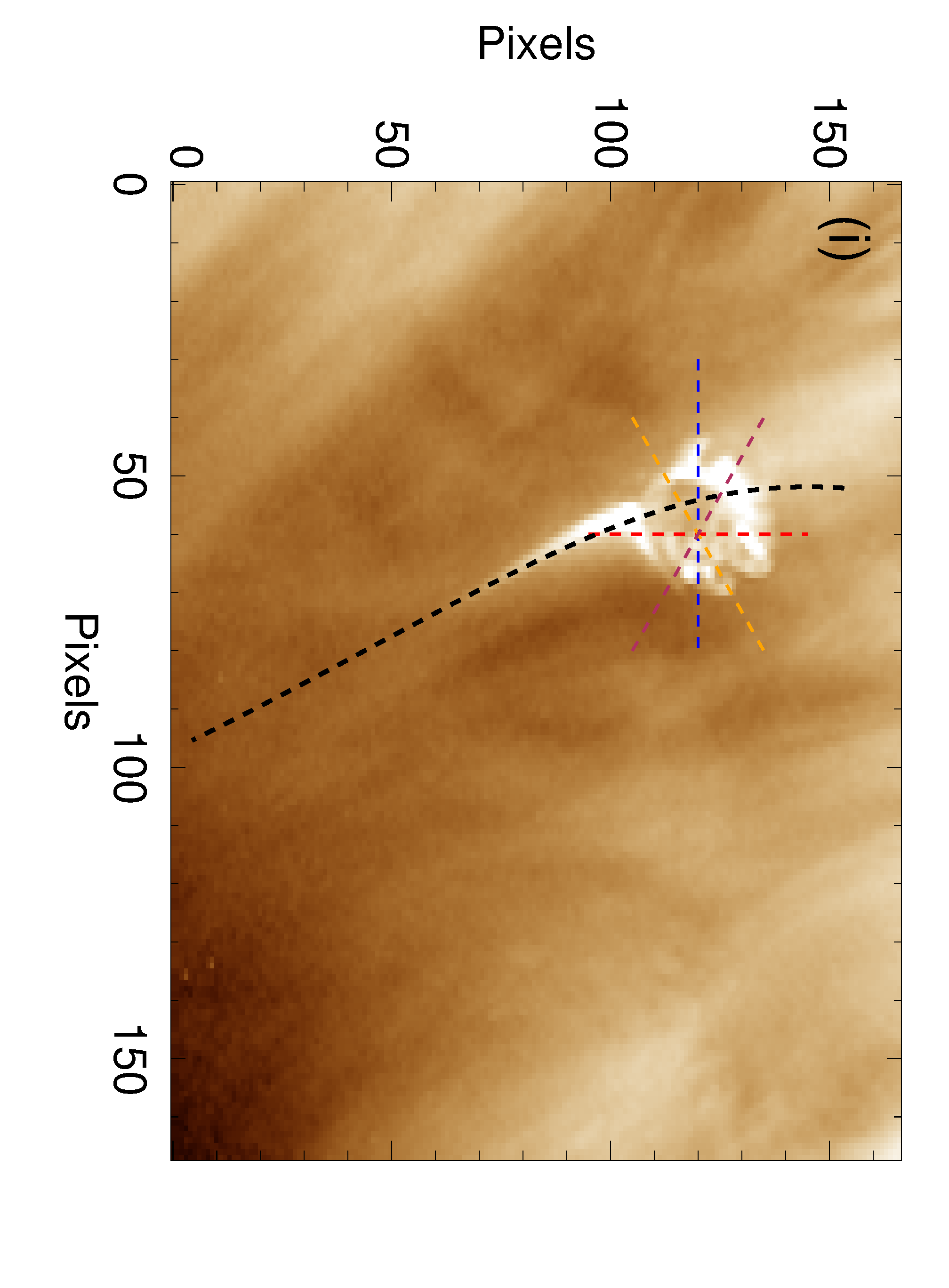}
\includegraphics[width=4.cm, height=3.cm]{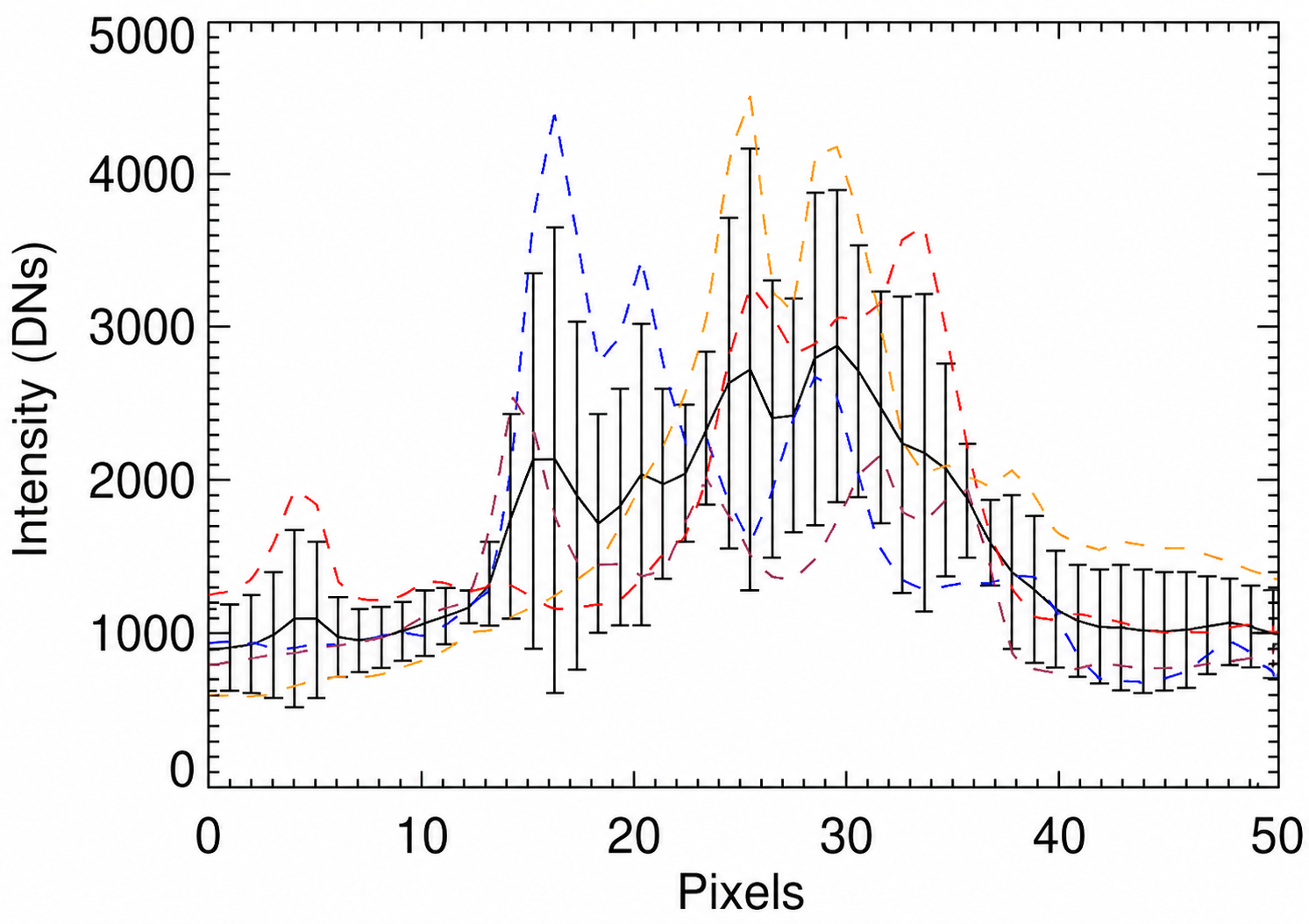}

\includegraphics[width=4.0cm, height=3.cm]{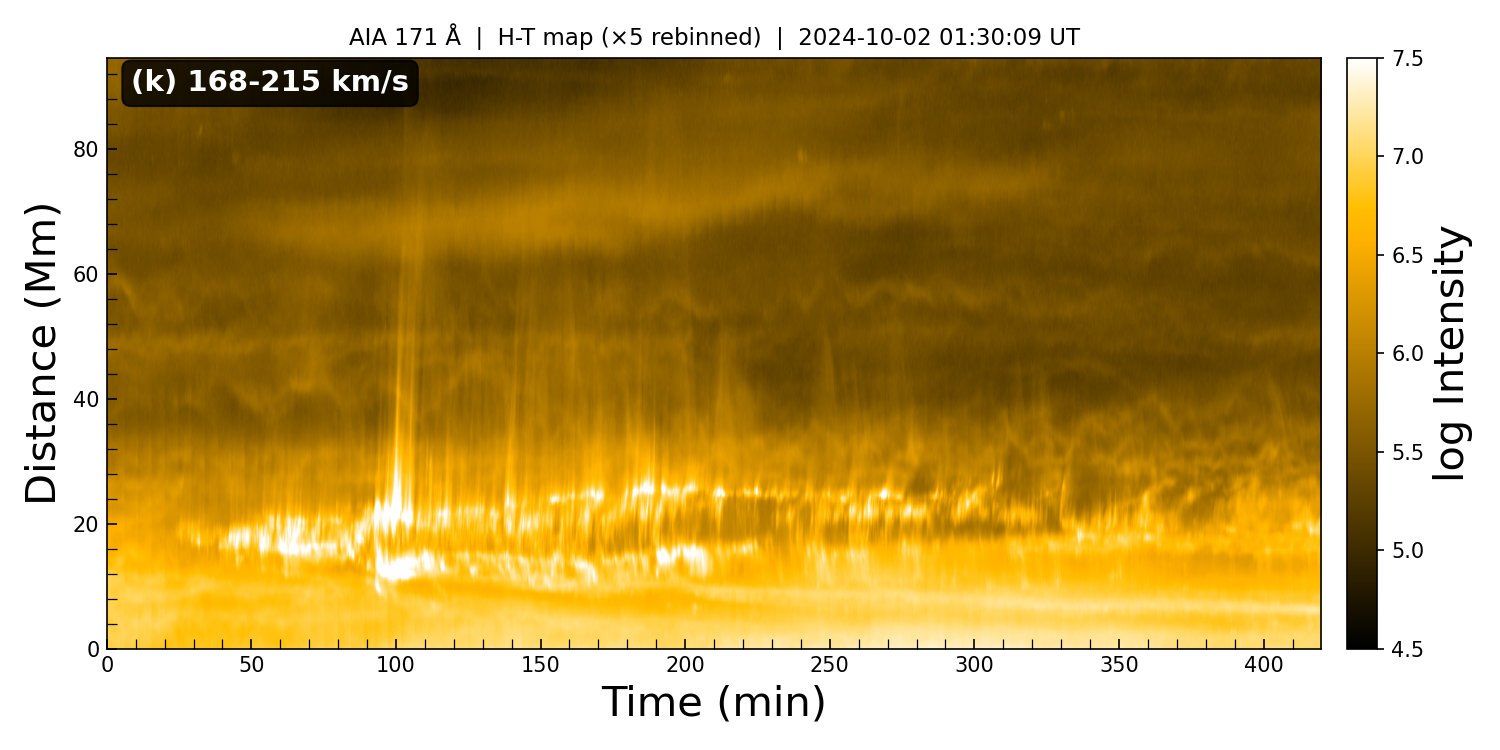}
}
\caption{Same as Figure~1 but for Jet~2 from a moderate-field-strength region of the Sun.}  
\label{fig:fig2}
\end{figure*}

 \subsection{Multi-wavelength Imaging Observations of Recurring Jets from Evolving Active Region of the Sun}
Jet~3 originates from an evolving active region with $\pm$200~G mixed-polarity fields and exhibits two distinct evolutionary phases captured in Figure~3: an initial anemone jet phase (panels a-d) and a subsequent recurring coronal jet eruptive phase (panels i-p; Figure~3). The magnetic topology during the eruptive phase (04:30--10:00~UT) has been analyzed in detail by \cite{2024ApJ...962L..38D}, while the complete triggering mechanism (01:00-11:00~UT), jet kinematics, thermal properties, cool-plasma ejections, steady localized heating, and the formation of hot ($>$5~MK) plasma near the jet base associated with two coronal nulls are presented by \cite{Mishra2026submitted}. Those studies show that successive flux emergence, extracted from the region outlined by the dashed box in panel~(a) of Figure~3, drives flux cancellation that produces multiple jets evolving from small-scale anemone jets into large-scale homologous coronal jets; persistent brightening in AIA~94~\AA\ ($T \sim 8$~MK) at the jet base further confirms a long-lived hot plasma component at the jet footpoints. Following the same hemispherical approximation adopted for Jets~1 and~2, loop lengths during both phases are estimated and listed in Table~1. The projected plane-of-sky speeds are 165-218~km~s$^{-1}$ during the anemone phase, increasing to 235-268~km~s$^{-1}$ during the three large-scale homologous coronal jets \citep{Mishra2026submitted}, consistent with more energetic reconnection driving the eruptive outflows.\\

\textit{Anemone jet phase.}
Panels (a)--(d) of Figure~3 show the source region at $t \approx$ 01:27~UT; AIA~171~\AA\ (panel~b) reveals a compact anemone jet with a bright base driven by fan-spine reconnection, while AIA~211 and 94~\AA\ (panels~c--d) confirm co-spatial hot emission indicating multi-thermal plasma ejection. Panel~(e) shows the normalized unsigned flux declining gently over $\sim$300~min (900~s running mean, red), and the detrended flux (panel~f) shows low-amplitude quasi-periodic fluctuations. The Morlet wavelet power spectrum (panel~g) reveals
statistically significant power at $\sim$4-18~min, and the global wavelet spectrum (panel~h) identifies a dominant period of $\sim$9.56~min ($a = 0.45$, 95\% red-noise significance), indicating that short-period quasi-periodic flux emergence drives the recurrent anemone jets.\\

\textit{Eruptive jet phase.}
Panels (i)-(l) capture the pre-eruptive fan-spine configuration at $t \approx$ 06:13~UT, with AIA~94~\AA\ (panel~l) showing compact hot emission at the reconnection site. Panels (m)-(p) show the large-scale eruptive jet at $t \approx$ 07:24~UT, visible as a broad collimated eruption in AIA~171 and 211~\AA\ (panels~n-o) with intense flare-like
hot emission in AIA~94~\AA\ (panel~p). Panel~(q) shows the unsigned flux declining steeply from $\sim$1.0 to $\sim$0.6 over 300-700~min (1800~s running mean), reflecting sustained pre-eruptive flux cancellation. The detrended flux (panel~r) shows larger-amplitude oscillations ($\sim$0.05-0.075) than the anemone phase, indicating more energetic photospheric driving. The wavelet power spectrum (panel~s) is dominated by long-period significant power at $\sim$32-64~min during 300-700~min, and the global wavelet spectrum (panel~t) identifies a dominant period of $\sim$32.15~min ($a = 0.76$, 95\% red-noise significance), the highest lag-1 autocorrelation among all three jets, reflecting the slow persistent flux cancellation characteristic of the pre-eruptive active region. The transition from $\sim$9.56~min (anemone phase) to $\sim$32.15~min (eruptive phase) within the same source region directly demonstrates how the evolving photospheric flux timescale governs the transition from recurrent confined jets to large-scale eruptions.\\

\begin{figure*}[ht]
\centering
\includegraphics[width=12.0cm, height=2.5cm]{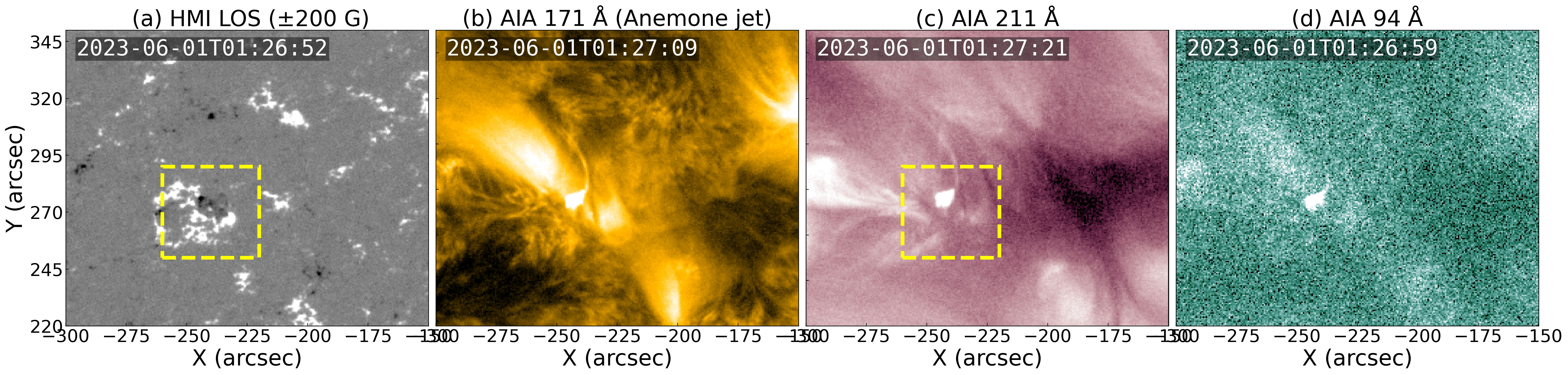}
\includegraphics[width=12.0cm, height=2.5cm]{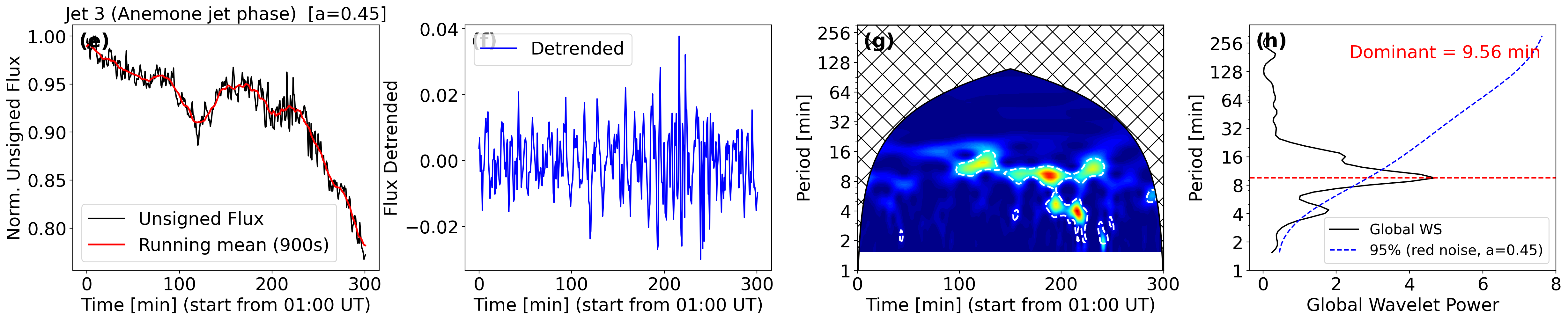} 
\includegraphics[width=12.0cm, height=2.5cm]{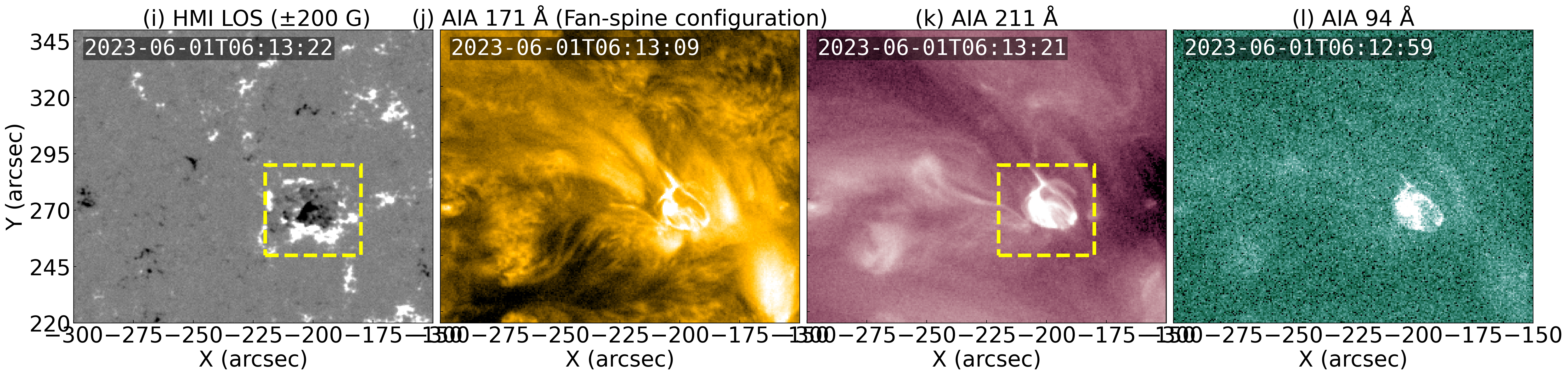}
\includegraphics[width=12.0cm, height=2.5cm]{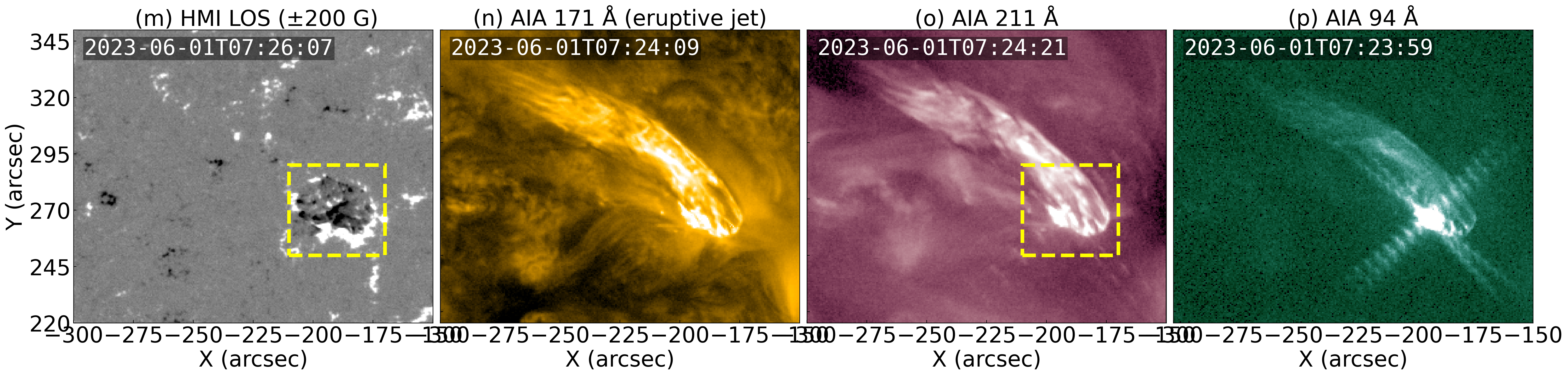}
\includegraphics[width=12.0cm, height=2.5cm]{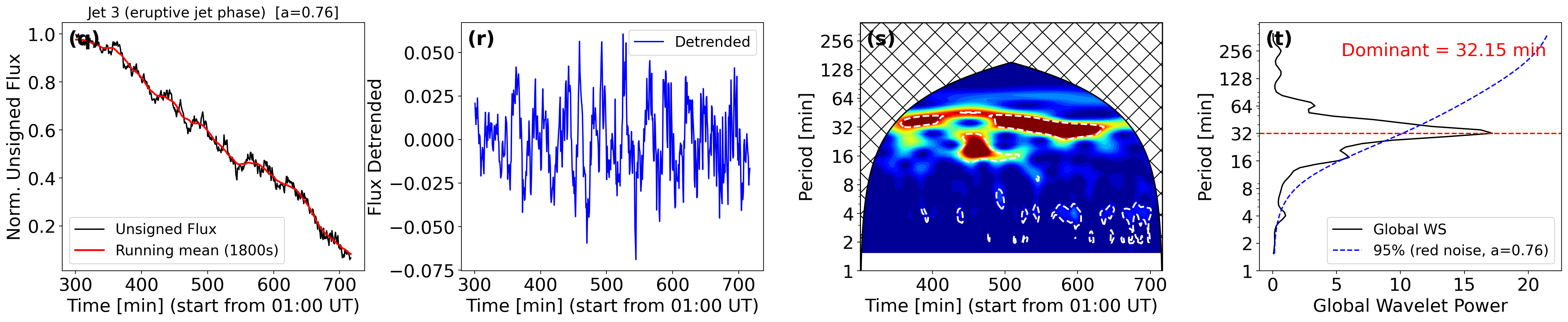} 
\caption{Same as Figure~1 but for Jet~3 from an evolving active region. Panels (a)-(h) correspond to the initial anemone jet phase (01:00-05:00~UT on 2023 June 1), when the null point lies low in the corona, with a dominant period of $\sim$9.56~min ($a = 0.45$). Panels (i)-(l) and (m)-(p) show the multi-thermal fan-spine configuration and the launch of large-scale coronal jets from the same ROI at $t \approx$ 06:13~UT and 07:24~UT, respectively. Panels (q)-(t) displayed the corresponding wavelet analysis during 05:00-12:00~UT, when the null point lies high in the corona, identifying a dominant period of $\sim$32.15~minutes ($a=0.76$), reflecting the slow pre-eruptive flux cancellation timescale.}
\end{figure*}
\begin{figure*}[ht]
\centering
\includegraphics[width=12.0cm, height=6.0cm]{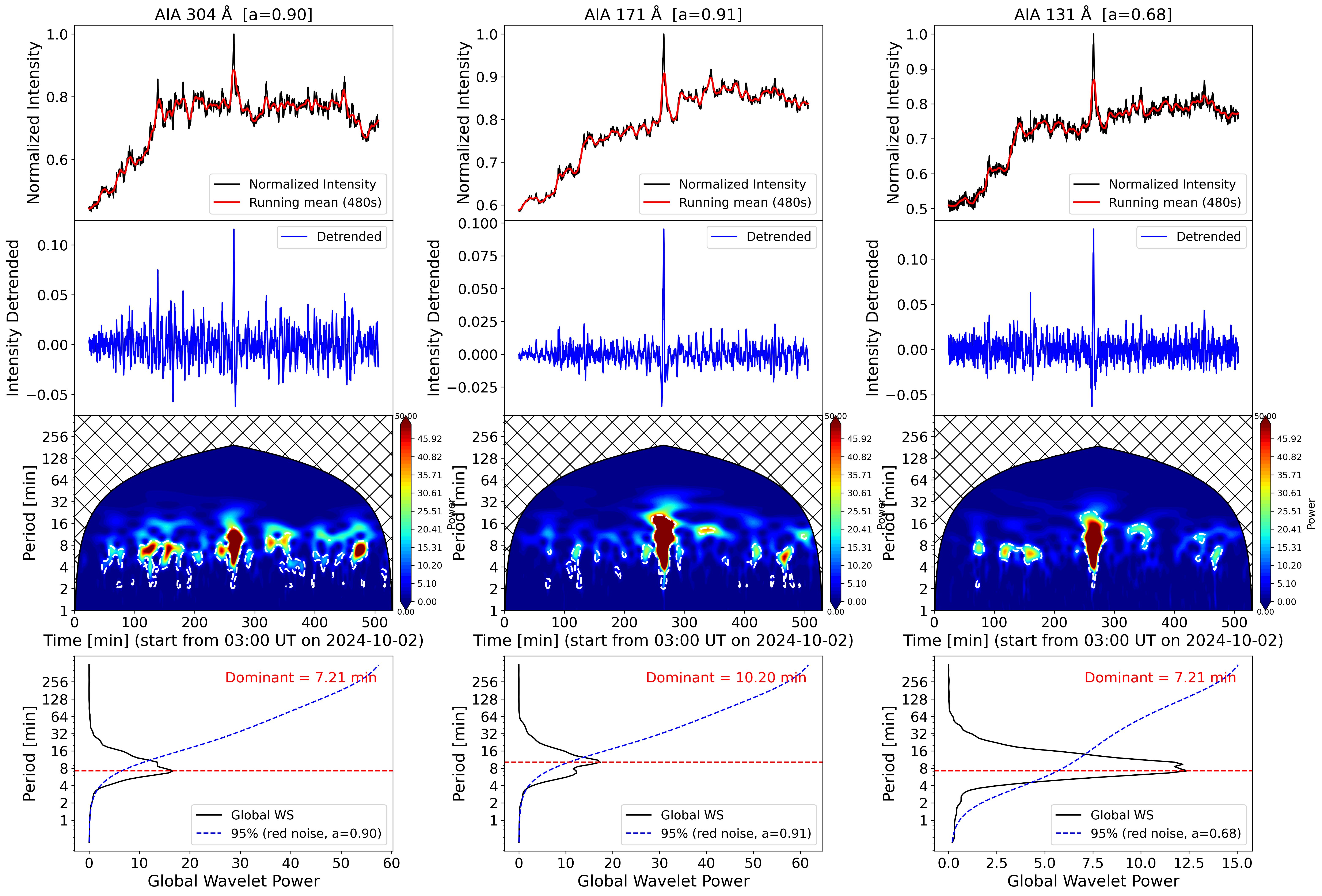}
\caption{Wavelet analysis for jet1 extracted from the light curve from the base of the anemone jets. The top panel shows the extracted light curve from the base of the jet, the second panel corresponds to the detrended light curve, the third panel shows the wavelet power maps, and the fourth panel shows the most prominent period for AIA 304, 171, and 131 {\AA}. }
\label{fig:fig3}
\end{figure*}
\begin{figure*}[ht]
\centering
\includegraphics[width=12.0cm, height=6.0cm]{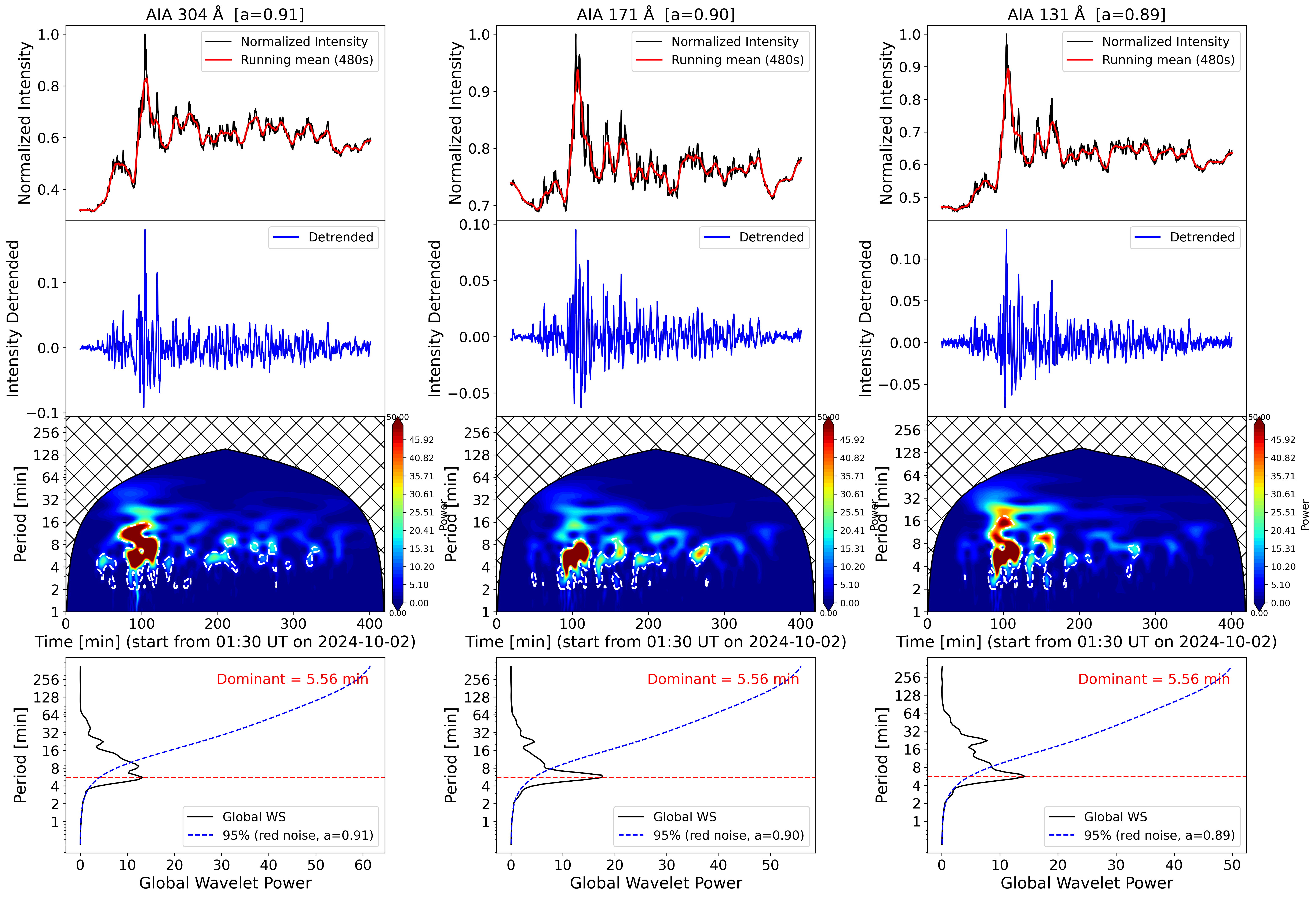}
\caption{Similar to Figure~4 but for Jet2.}  
\label{fig:fig4}
\end{figure*}

\begin{table*}
\begin{center}
\begin{minipage}{\textwidth}
\caption{Different physical and geometrical parameters were derived for all three studied jets. A detailed analysis of Jet~3 is presented in \cite[submitted][]{Mishra2026submitted}. The background electron number density is estimated to be 3-5$\approx 10^{9}\,\mathrm{cm^{-3}}$. The HMI magnetic flux exhibits a periodicity of about 8-12 minutes, strongly supporting the interpretation of periodically modulated magnetic reconnection associated with flux emergence. }\label{tab5}
\scriptsize
\begin{tabular*}{\textwidth}{@{\extracolsep{\fill}}lccccccc@{\extracolsep{\fill}}}
\hline

Events & First Appearance  & $\sim$Loop Length & Periods (cool) & Periods (Hot) & $\rho_{base}-\rho_{back.}$ &Avg. T (MK)\\
\hline
\hline
Jet1	&	2024-10-02 03:20  & 6-7 Mm &	6-9  &	6-8 (Not in AIA 94)	&	3.5-3.9$\times$10$^{9}$  & 1.95-2.05 \\
Jet2	&	2024-10-02 01:52  & 6-8 Mm  	 &	6-9	& 5-9	&	7.5-9.8 $\times$10$^{9}$ & 2.0-2.2\\
Jet3 (Initial)	& 2023-06-01 01:05	&4-6 Mm	&	6-8	&	7-9	&	2.1-2.9 $\times$10$^{10}$  & 1.7-3.2 \\
Jet3 (Eruptive)	&	2023-06-01 04:38&6-10 Mm&	12-13	&	11-13	&	3.2-6.5 $\times$10$^{10}$  & 1.8-5.4 \\
   
\hline

\end{tabular*}
\end{minipage}
\end{center}
\end{table*}

\subsection{Wavelet Analysis and Periodic Magnetic Reconnection}
\begin{figure*}[ht]
\centering{}
\includegraphics[width=12.0cm, height=6.0cm]{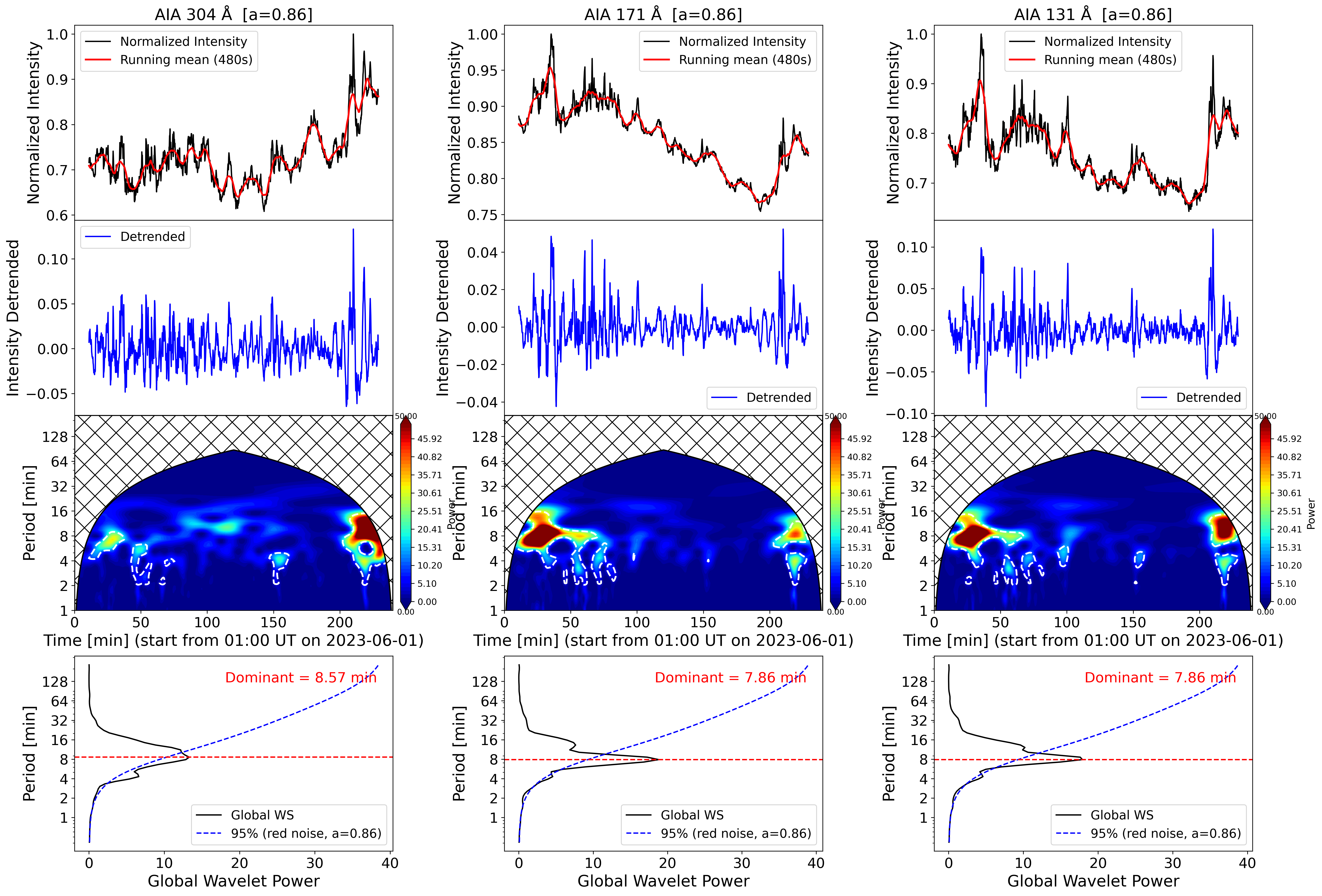}
\caption{Similar to Figure~4 but for Jet3 during its initial anemone jet phase (01:00 and 05:00 UT on June 1,$^{st}$ 2023), when the null point lies low in the solar corona.}  
\label{fig:fig5}
\end{figure*}

\begin{figure*}[ht]
\centering
\includegraphics[width=12.0cm, height=6.0cm]{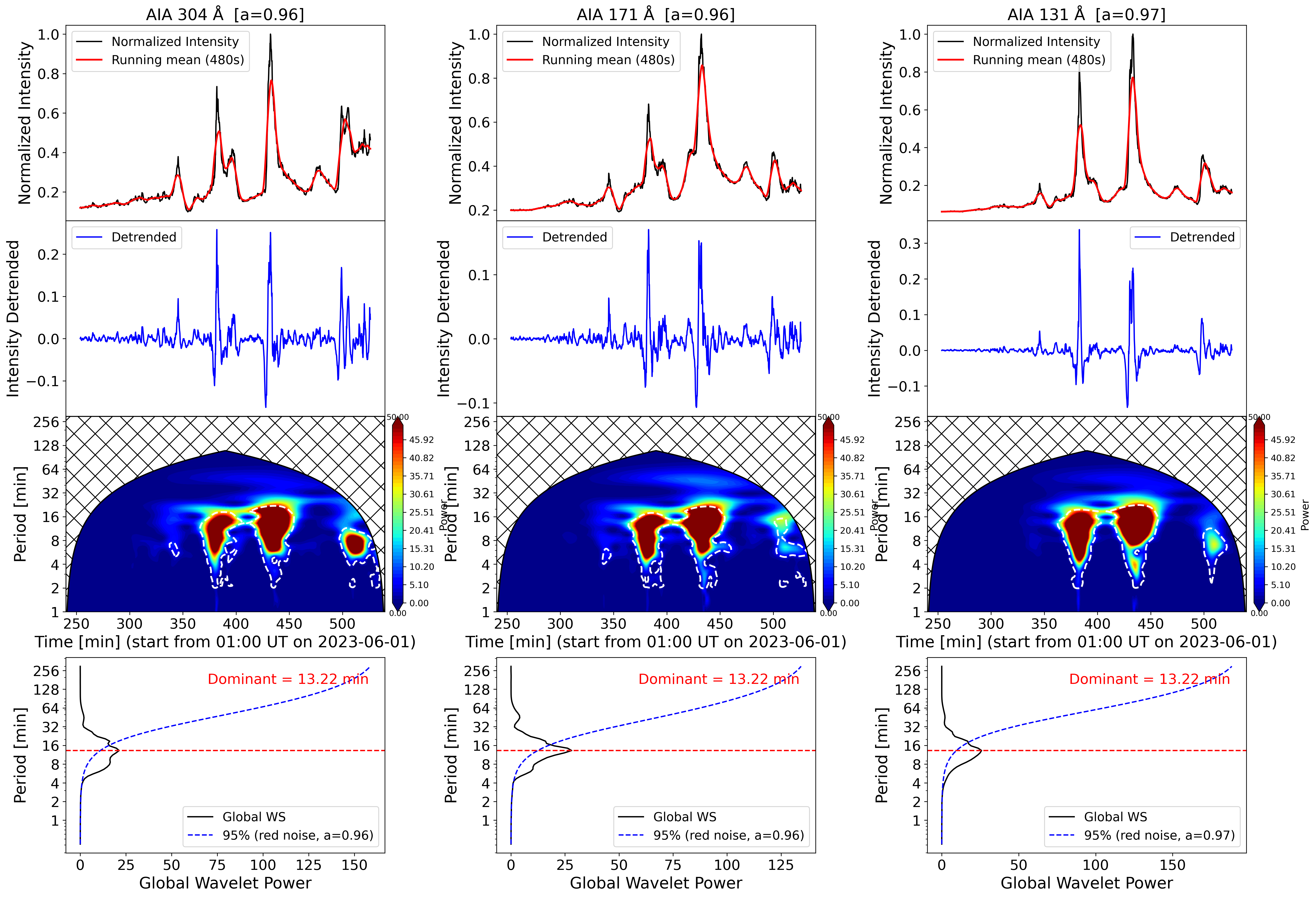}
\caption{Similar to Figure~4 but for Jet3 during its eruptive phase (05:00-10:30 UT on June, 1$^{st}$, 2023), when the null point lies high in the solar corona and launches three homologous coronal jets.}  
\label{fig:fig6}
\end{figure*}

To investigate the periodic intensification in the kinematic profiles, we extracted light curves from the jet bases across all AIA EUV channels. Figures~\ref{fig:fig3}-\ref{fig:fig6} present the wavelet power spectra and global wavelet profiles for three jets observed in different magnetic environments. The top row of each figure displays the temporal evolution of the normalized intensities at the jet bases. The second row shows the corresponding detrended light curves obtained after removing long-term trends. The third row presents the wavelet power spectra, highlighting regions of significant power in time-period space, while the fourth row represents the global wavelet spectrum (period versus power) for each case. The first, second, and third columns correspond to the AIA 304~\AA, 171~\AA, and 131~\AA\ passbands, respectively.

Figure~\ref{fig:fig3} shows the wavelet analysis for Jet~1, emerging at the quiet-sun region. The wavelet power maps reveal significant periodicities in the range of $P \approx 7$-$ 10$ minutes across all channels, with the strongest power localized near the onset of recurrent intensity enhancements. The persistence of these periods across multiple cycles suggests that the observed QPPs are not random fluctuations but are instead associated with repetitive energy release events. Fig.~\ref{fig:fig4} presents the wavelet analysis for Jet~2, associated with the moderate-field region, which exhibits dominant periods of $P \approx 6$ minutes. The enhanced power in the hot coronal channel (131~\AA) suggests that periodic magnetic reconnection at the jet base produces multi-thermal plasma, including components exceeding 8-10~MK (periodicity for AIA 94 {\AA} is $\approx$9-11 minutes, which formed at $\sim$8 MK ). The temporal alignment of these periodicities with flux emergence episodes supports the interpretation that magnetic reconnection is the primary driver of these jets.
Figs.~\ref{fig:fig5} and \ref{fig:fig6} for Jet~3, which shows a moderate field strength region, exhibit two distinct phases. During its initial evolution with a low-lying null point (Figure~\ref{fig:fig5}), the dominant periods are $P \approx 7$-$9$~minutes. In the later phase, when the null point rises and launches three homologous coronal jets (Figure~\ref{fig:fig6}), the periods extend to $P \approx$-$13$~minutes. The observed increase in periodicity correlates with the expansion of the magnetic structure and the development of large-scale jets, consistent with reconnection models in which the global magnetic geometry controls the characteristic reconnection timescale \cite{RevModPhys.82.603}.
The extracted QPP periods ($6$--$13$~minutes) are significantly longer than typical chromospheric oscillations ($\sim 3$~min) or p-mode oscillations ($\sim 5$min), ruling out wave leakage as the dominant mechanism. Instead, these results support periodic magnetic reconnection within a fan-spine-like magnetic configuration in the presence of successive flux emergence as the origin of QPPs \cite{2012A&A...548A..98M,2017ApJ...844....2T}. We would like to highlight that the quasi-periodic ridges in the AIA~171~\AA\ distance-time maps for all three jets studied here are co-spatial with episodic multi-thermal brightenings at the jet base observed across AIA~304, 171, and 131~\AA\ (Figures~1-3), which is consistent with periodic reconnection-driven plasma ejections within the fan-spine null-point configuration. The SDO/HMI unsigned flux at the jet source regions yields dominant periodicities of $\sim$10-32~min (Figures~1-3), whereas the corresponding coronal AIA periodicities are $\sim$5-13~min (Figures~4-7), roughly a factor of $\sim$2 shorter. This approximate harmonic relationship suggests that each photospheric flux emergence episode may trigger multiple reconnection bursts, plausibly through current-sheet fragmentation or plasmoid instability \citep{2006A&A...452..343N}, whereby a single flux emergence event drives two or more successive plasma ejections on shorter coronal timescales. At the photospheric level, the magnetic field structures are relatively smooth and strongly coupled, whereas the coronal responses are more dynamic. Within a single episode of flux emergence, the same emerging system can drive multiple episodes of periodic magnetic reconnection, which may explain why shorter characteristic periods are detected in the coronal AIA channels. We note that the morphological resemblance of the time-distance ridges to propagating slow magnetoacoustic wave signatures \citep{2024MNRAS.527.5302M} cannot be entirely excluded, since slow waves can also be excited by the reconnection process itself. However, the close correspondence between photospheric flux periodicities and coronal EUV intensity modulations, together with the absence of clear shock or wave-mode signatures, collectively favors reconnection-driven periodic plasma ejections as the primary interpretation.\\

In conclusion, wavelet analysis reveals that QPPs with periods of 6-13 minutes are persistently detected in three jets across all EUV channels, which appear at different spatiotemporal scales in the presence of a fan-spine topology. The extracted periodicities from the base of the jets, across different EUV channels, coincide with the phases of magnetic flux emergence and cancellation, indicating that the emerging flux reconnects periodically or quasi-periodically and drives the observed intensity modulations. The presence of multithermal plasma, including recurrent hot components ($>5~\mathrm{MK}$), implies that periodic reconnection contributes to localized coronal heating even in quiet and moderate field strength regions of the Sun. Together with previous studies on QPPs and coronal energy transport \cite{2009SSRv..149..119N,2020STP.....6a...3K,2021SSRv..217...66Z}, these results underscore the crucial role of small-scale magnetic activity in regulating mass and energy circulation in the localized solar corona.

\subsection{Thermal Properties of the Anemone Jet}
We performed DEM analysis to investigate multi-thermal plasma, heating, and cooling processes, as well as the role of quasi-periodic magnetic reconnection in driving recurring anemone and large-scale coronal jets. DEM-weighted temperature and EM maps at $t = 07{:}25$ UT (Jet1) and $t = 03{:}10$ UT (Jet2) reveal coronal temperatures in the extended jet spires with cool plasma components in the AIA 304~\AA\ channel (Figure~1, 2, and 3). We want to highlight that the jet bases are dominated by coronal and hot plasma exceeding 5 MK episodically (panels a, b, c, and d; Figure~8), including regions exceeding it $10~\mathrm{MK}$ for Jet3 when the null lies high in the solar corona \cite{2024ApJ...962L..38D} and \cite[submitted][]{Mishra2026submitted}, indicating intense, localized heating from magnetic reconnection \cite{2023NatCo..14.2107C}.

Panels (e), (f), (g), and (h) of Figure~8 show the temporal evolution of EM and EM-based average temperature extracted from the jet base for Jet1 and Jet2, respectively, while detailed thermal and EM analysis for Jet3 is found in \cite[submitted][]{Mishra2026submitted}. The temporal evolution of EM maps shows quasi-periodic modulations during the launches of the jets, while the temporal evolution of EM-based average temperature shows significant episodic enhancement, as both cool and hot plasma formed after the launch of each jet, indicating recurrent energy release and repeated multi-thermal plasma formation consistent with oscillatory reconnection \cite{2025ApJ...986..197H}. We found that pronounced EM increases occur during significant jet episodes, with sharp temperature spikes at onset, followed by cooling and an increasing contribution from cooler components (panels e and f; Figure~7). This spatiotemporal evolution illustrates the formation of persistent hot plasma above the typical solar corona at jet sites, which supports steady localized heating near the reconnection region \cite{2023NatCo..14.2107C}.
The thermal energy at the jet base is
\(E_{\mathrm{th}} = \tfrac{3}{2}\, n\, k_{\mathrm{B}}\, T\, V\),
where \(n\) is the electron number density, \(k_{\mathrm{B}}\) is Boltzmann's constant, \(T\) is the DEM-weighted temperature, and \(V\) is the jet-base volume. Assuming a hemispherical base, the volume is
\(V = \tfrac{2}{3}\,\pi\,R^{3}\),
with \(R\) the characteristic base radius. We use the different physical parameters from Table~1 to extract the thermal energies, which is $10^{26}\text{--}10^{27}~\mathrm{erg}$ (Jet 1), $10^{27}\text{--}10^{28}~\mathrm{erg}$ (Jet 2), and $10^{27}\text{--}10^{29}~\mathrm{erg}$ (Jet 3). The estimated thermal energy significantly exceeds typical CBP energies ($10^{23}\text{--}10^{24}~\mathrm{erg}$) \cite{1975SoPh...41..381G, 2021ApJ...906...59H} for Jet1 and Jet2 but is comparable to network flares associated with CBPs ($\sim 10^{26}\text{--}10^{27}~\mathrm{erg}$) \cite{2002AdSpR..30..647P} and magnetic-reconnection-driven CBP events ($\sim 10^{27}\text{--}10^{28}~\mathrm{erg}$) \cite{2019LRSP...16....2M}. The estimated thermal energies represent upper limits, as we adopt a filled hemispherical geometry with a unity filling factor ($f = 1$) following \cite{2019LRSP...16....2M}. A hemispherical shell confined between two radial distances would reduce the emitting volume and consequently lower the number density and thermal energy; however, since the LOS depth cannot be constrained from AIA imaging alone, the filled hemisphere is retained as a first-order approximation. These results indicate that the studied anemone jets represent energetic CBP-associated events, in which recurrent, periodic magnetic reconnection efficiently converts magnetic energy into thermal energy sufficient to contribute to the localized coronal heating.
\begin{figure*}
\includegraphics[width=12.0cm, height=8.0cm]{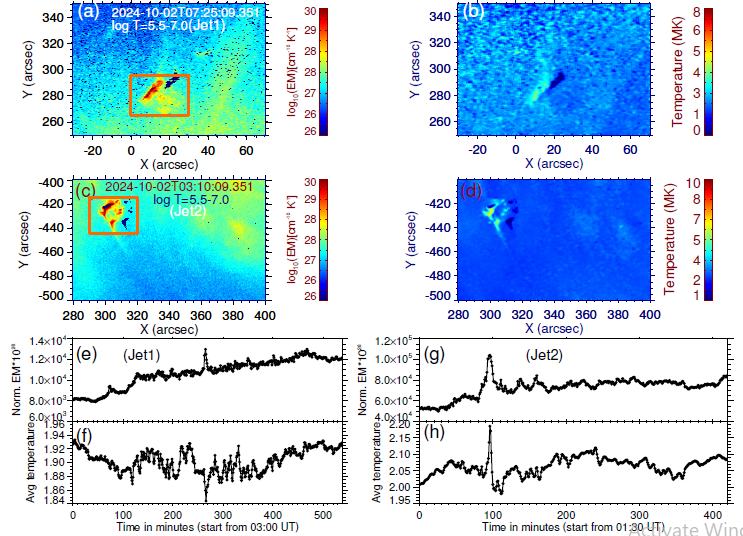}
\caption{Top panel: Panels (a) and (b) depict the EM map and EM weighted-temperature map for Jet~1 at t=07:25 UT. Middle panel: As in the top panels, EM and EM-weighted temperature for Jet2 at t=03:10 UT. Bottom panels: Panels (e), (f), (g), and (h) show the temporal evolution of EM for different temperature bins and the weighted average temperature extracted from two boxes from the base of both jets, respectively}.
\label{fig:fig7}
\end{figure*}
\subsection{Cooling Timescales and Repetitive Reconnection-Driven Heating in Localized Quiet-Sun, Moderate-Field-Strength, and Active  Coronal Regions}

In this paper, three long-duration recurring anemone and coronal jets occurring in quiet Sun, moderate-field-strength, and evolving AR regions are examined. These events are associated with the presence of long-lived fan-spine magnetic topology. To investigate heating and cooling mechanisms, the radiative and conductive cooling timescales near the jet footpoint are evaluated. On average temporal scales, if the radiative cooling time exceeds the reconnection periodicity ($\tau_{cool}$ $\geq$ T$_{period}$), the plasma can be reheated before it thoroughly cools down near the base of the jet, maintaining elevated temperatures and constraining the thermodynamic evolution (Figure~7).

The radiative cooling time is, $\tau_{\mathrm{rad}} \approx 5\times 10^{3} \left(\frac{T}{10^{6}\,\mathrm{K}}\right)^{3/2} \left(\frac{n}{10^{9}\,\mathrm{cm}^{-3}}\right)^{-1}\,\mathrm{s}$ \cite{2011LRSP....8....6S}, with the radiative loss function approximated as $Q(T) \approx 10^{-22} \left(\frac{T}{10^{6}\,\mathrm{K}}\right)^{-1/2}$ for $T < 10^{7}\,\mathrm{K}$ and $Q(T) \approx 3\times 10^{-23} \left(\frac{T}{10^{7}\,\mathrm{K}}\right)^{1/2}$ for $T > 10^{7}\,\mathrm{K}$ \cite{2011LRSP....8....6S}. The conductive cooling time is $\tau_{\mathrm{cond}} \approx \frac{3\, n_{e} k_{\mathrm{B}} L^{2}}{\kappa_{0} T^{5/2}}$ with $\kappa_{0} = 9.2\times10^{-7}\,\mathrm{erg\,s^{-1}\,cm^{-1}\,K^{-7/2}}$ \cite{2015RSPTA.37340256K,2011LRSP....8....6S}.

For the CBPs associated with the jet base, they may be treated as dense and semi-spherical magnetic structures. The extracted different physical and geometrical parameters near the jet base are mentioned in Table~1. Using the loop length, number density from the base of the jets, and DEM-weighted average temperature, the estimated radiative cooling time for Jet1 is 20-30 minutes, for Jet2 is 10-20 minutes, and for Jet3 (initial and eruptive phases) is 5-15 minutes, respectively. Similarly, the extracted conductive cooling times range from 3 to 8 minutes for all studied jets. Thus, the radiative cooling timescale ($\tau_{\mathrm{rad}}$) exceeds both the conductive cooling timescale and the dominant quasi-periodic pulsation period identified from multiple AIA EUV channels via wavelet analysis, implying that the plasma cannot cool efficiently between successive heating events and suggesting persistent, quasi-continuous heating in quiet-Sun and moderate-field regions \cite{2015RSPTA.37340256K}. Long-lived anemone jets undergo sequential evolution in which compact bright points form and host significantly hotter plasma than the surrounding quiet corona, potentially contributing to quiet-Sun heating \cite{2019LRSP...16....2M}. For all three jets, the conductive cooling times are of the order of 3-8 minutes, while the radiative cooling times are significantly longer ($\sim$10–30 minutes). It means that, on average, over time, during a reconnection period of 6-13 minutes, the plasma typically does not fully cool radiatively between reconnection bursts. Therefore, the repeated reconnection can maintain elevated heated loops and a quasi-steady, multi-thermal structure at the jet bases, which can contribute to the localized coronal heating.

\section{Discussion \& Conclusions}
\label{sec: conclusion}
 In this study, we explore the presence of QPPs in three long-lived coronal jets exhibiting a fan-spine-like magnetic configuration in the quiet Sun and regions of moderate field strength. MHD oscillations, such as kink, sausage, and torsional Alfvén waves, and reconnection-driven processes such as oscillatory reconnection, plasmoid instability, and intermittent flux emergence are considered to be the key mechanisms to trigger the QPPs in solar eruptions \cite{2009SSRv..149..119N, 2018SSRv..214...45M, 2021SSRv..217...66Z}. In the context of MHD wave-driven QPPs, \cite{2022MNRAS.511.4134S} showed that Alfv\'{e}n pulse-driven magnetoacoustic shocks produce quasiperiodic jets with 
periods of $\sim$4~minutes. Also, the reconnection-driven QPPs with periodicity of $\sim$45~s and $\sim$3~minutes have been reported in recurrent and kink-unstable jets, respectively \cite{2022FrASS...932099L, 2023ApJ...945..113M}. Recurrent jets with regular periodicities  \cite{2012A&A...542A..70M, 2014A&A...561A.134Z,2022ApJ...933...21K}, further establishing reconnection-driven periodicity as a common feature of jet-associated QPPs. Recently, \cite{Gu_2026} proposed that quasiperiodic magnetic reconnection, modulated by slow magnetoacoustic waves in the chromosphere, links wave dynamics directly to the periodicity of jet eruptions.\\
In this paper, we find that successive flux emergence episodes in the fan-spine configuration periodically drive reconnection with the overlying field, triggering QPPs with periods of 6-13 minutes. These periods exceed those reported in active-region jets \cite{2022FrASS...932099L, 2023ApJ...945..113M}, Alfvén-driven jets \cite{2022MNRAS.511.4134S}, p-modes ($\sim$5~minutes), and chromospheric oscillations ($\sim$3~minutes) \cite{2021SSRv..217...66Z}, suggesting that the lower reconnection rate in quiet-Sun fan-spine configurations governs the longer periodicity and rules out MHD wave mechanisms as primary drivers. The repetitive formation of multi-thermal plasma serves as an indirect signature of periodic reconnection \citep{2022FrASS...932099L, 2023ApJ...945..113M, Gu_2026}, and the QPP periods exceeding both radiative and conductive cooling times \cite{2011LRSP....8....6S} indicate that hot plasma is steadily maintained near the jet base throughout each event, further supporting periodic reconnection as the dominant driver.

Three jets studied here occur within a fan-spine-like magnetic reconnection configuration that favors recurrent jet-like eruptions at different field strengths in the solar atmosphere. The estimated multi-thermal periodicity from the wavelet analysis is $6$-$13$ minutes, consistent with quasi-periodic oscillations reported in CBPs/XBPs, with periods ranging from $30$ seconds to $60$ minutes \cite{2019LRSP...16....2M}. Our wavelet and magnetogram analysis confirm that periodic magnetic reconnection is the primary driver of these jets. The rate of magnetic flux emergence and the height of the coronal null point control the thermal response and persistence of hot plasma ($>5~\mathrm{MK}$) at jet bases, which is typically higher than the reported average temperature of CBPs \cite{2008A&A...485..289K, 2013ApJ...768...32C}. Over timescales of several hours, the persistent presence of such hot plasma implies steady or quasi-steady localized coronal heating even in the quiet and moderate-field strength of the solar atmosphere and supplements the presence of hot plasma in the presence of fan-spine topology for ARs \cite{2023NatCo..14.2107C, 2024NatAs...8..706L}, with estimated thermal energies $10^{26}\text{--}10^{29}~\mathrm{erg}$ depending on the magnetic environment and jet properties \cite[submitted][]{Mishra2026submitted}. Therefore, periodic magnetic reconnection within fan-spine-like configurations is an efficient and persistent mechanism for both plasma acceleration and localized energy deposition, thereby contributing significantly to the heating of the localized chromosphere and corona in small-scale magnetic structures. The reconnection-driven heating identified here is confined to a localized fan-spine null-point configuration and cannot be considered a driver of global coronal heating \cite{1996PASJ...48..353Y, 2015RSPTA.37340256K, 2019ApJ...887..137S, 2025ApJ...982..147M, 2025ApJ...984...36S}. Nevertheless, null-point fan-spine topologies are ubiquitous in the quiet-Sun corona \cite{2023NatCo..14.2107C}, and the cumulative effect of numerous such small-scale, episodic reconnection events, each individually localized, may collectively contribute to sustaining the thermal structure of the low corona, consistent with the nanoflare storm scenario \cite{1988ApJ...330..474P, 2018NatCo...9..692Y}.\\

In conclusion, our results highlight the importance of such reconnection geometry in regulating the QPPs and launches of small-to-large-scale jets. The long-lived jets associated with CBPs in the presence of a fan-spine-like configuration and successive flux emergence undergo recurrent magnetic reconnection and maintain hot plasma at their bases, thereby contributing to localized coronal heating across the quiet and moderate-field strengths of the Sun. Successive flux emergence in the presence of such geometry regulates jet occurrence and quasi-periodic pulsations (QPPs) through periodic reconnection and MHD wave modulation \cite{2014Sci...346A.315T, 2015ApJ...806..172S}. The scale-dependent periodicity of QPPs found in our study also aligns with the statistical trends reported by Hayes et al. (2020) \cite{2020ApJ...895...50H}. The finding of Hayes et al. (2020) \cite{2020ApJ...895...50H} suggests that the rate of reconnection and the geometry of the loop are the critical key factors. Similar to \cite{2023NatCo..14.2107C,2024NatAs...8..706L} for active regions, the present study highlights the importance of fan-spine configurations, together with successive flux emergence, in triggering recurring jet activity and associated QPPs. It demonstrates that QPPs play a crucial role in energy transport and localized coronal heating in the quiet Sun and moderate-field-strength regions of the solar atmosphere \cite[submitted][]{Mishra2026submitted}. Future high-resolution observations from Solar-C, DKIST, Solar Orbiter, and Aditya-L1, combined with detailed spectroscopic observations and 3D MHD simulations, are necessary for understanding the energy and mass circulations in the localized quiet and moderate-field regions of the solar atmosphere. A larger statistical sample is required to quantify how fan-spine-like reconnection geometries (or null-point-type configurations) facilitate periodic reconnection and how the resulting repeated energy release contributes to localized coronal heating. \\

\ack{We thank the anonymous reviewers for their constructive comments and valuable suggestions that helped to improve this manuscript. S.K. Mishra acknowledges Kyoto University, Japan, for providing fellowship support and access to computational facilities. Balveer Singh acknowledges the project titled “100 years of the global solar corona: new insights from historical data”, which is funded by the Science and Technology Facilities Council (STFC), United Kingdom Research and Innovation(UKRI) through grant ST/W001098/1. K.S. and D.Y. were supported by the National Natural Science Foundation of China (NSFC,12173012, 12473050), the Guangdong Natural Science Funds for Distinguished Young Scholars (2023B1515020049), the Shenzhen Science and Technology Project (JCYJ20240813104805008), the Shenzhen Key Laboratory Launching Project (No. ZDSYS20210702140800001), and the Specialized Research Fund for State Key Laboratory of Solar Activity and Space Weather. We thank Torrence and Compo (1998) for publicly making their wavelet analysis tools available. Data are courtesy of NASA/SDO and the AIA science team. }
\begin{fmtext}
\end{fmtext}




\bibliographystyle{vancouver}
\bibliography{sample}

\end{document}